\documentclass[]{aa} 
\usepackage{graphicx}            
\usepackage{epsfig}
\usepackage{epstopdf}
\usepackage{txfonts}
\usepackage{natbib}
\usepackage{gensymb}

\newcommand{\raiseChi}

\begin{document}

   \title{High resolution observations of the outer disk around T Cha: the view from ALMA}


\author{N. Hu\'elamo\inst{1}
          \and
          I. de Gregorio-Monsalvo\inst{2,3}
          \and
          E. Macias\inst{4}
          \and
          C. Pinte\inst{5,6}
          \and
          M. Ireland\inst{7}
          \and
          P. Tuthill\inst{8}
          \and
         S. Lacour\inst{9}
                    }
  \institute{Centro de Astrobiolog\'{\i}a (INTA-CSIC);  ESAC Campus, P.O. Box 78, E-28691 Villanueva de la Ca\~nada, Spain\\
         \email{nhuelamo@cab.inta-csic.es}
         \and
         Joint ALMA Observatory (JAO), Alonso de C\'ordova 3107, Vitacura, Santiago de Chile
         \and
         European Southern Observatory, Garching bei M\"unchen, D-85748 Germany
         \and
          Instituto de Astrof\'{\i}sica de Andaluc\'{\i}a, CSIC,  Glorieta de la Astronom\'{\i}a s/n, E-18008 Granada, Spain                
          \and
          UMI-FCA, CNRS/INSU France (UMI 3386), and Dpto. de Astronom\'{\i}a, Universidad de Chile, Casilla 36-D Santiago, Chile
          \and
          Univ. Grenoble Alpes, IPAG, 38000 Grenoble, France; CNRS, IPAG, 38000 Grenoble, France
          \and 
          Research School of Astronomy and Astrophysics, Australian National University, Canberra ACT 2611, Australia
          \and
         Sydney Institute for Astronomy, School of Physics, University of Sydney, NSW 2006, Australia
         \and 
          LESIA, CNRS\/UMR-8109, Observatoire de Paris, UPMC, Universit\'e Paris Diderot, 5 place Jules Janssen, 92195 Meudon, France
             }
   \date{Received ---,---; accepted ----, ----}
  \abstract
   {Transitional disks are circumstellar disks with dust gaps thought to be related in some cases with planet formation. They can shed light on the planet formation process by the analysis of their gas and dust properties. T Cha is a young star surrounded by a transitional disk with signatures of planet formation}
 {The aim of this work is to study the outer disk around T~Cha  and to derive its main properties.}
   {We have obtained high-resolution and high-sensitivity ALMA observations in the ${\rm CO}(3$--$2)$,  ${\rm ^{13}CO}(3$--$2)$, and ${\rm CS}(7$--$6)$ emission lines to reveal the spatial distribution of the gaseous disk around the star. In order to study the dust within the disk we have also obtained continuum images at 850 $\mu$m from the line-free channels.}
   {We have spatially resolved the outer disk around T~Cha. Using the CO(3-2) emission we derive a radius of $\sim$230\,AU.  We also report the detection of the 
   $^{13}$CO(3-2) and the CS(7-8) molecular emissions, which show smaller radii than the CO(3-2) detection.  The continuum observations at 850\,$\mu$m  allow the spatial resolution of the dusty disk, which  shows two emission bumps separated by $\sim$40\,AU, consistent with the presence of a dust gap in the inner regions of the disk, and an outer radius of $\sim$80\,AU.  Therefore, T~Cha is surrounded by a compact dusty disk and a larger and more diffuse gaseous disk, as previously observed in other young stars. The continuum intensity profiles are different at both sides of the disk suggesting possible dust asymmetries. 
   We derive an inclination of  $i (\degree)$\,=\,67$\pm$5, and a position angle of $PA (\degree)$\,=\,113$\pm$6, for both the gas and dust disks. The comparison of the ALMA data with radiative transfer models shows that the gas and dust components can only be simultaneously reproduced when we include a tapered edge prescription for the surface density profile. The best model suggests that most of the disk mass is placed within  a radius  of   $R<$ 50\,AU.
   Finally, we derive a dynamical mass for the central object of  $M_{*}$=1.5$\pm$0.2\,M$_{\odot}$, 
   comparable to the one estimated  with evolutionary models for an age of $\sim$10\,Myr.}
  {}
  \keywords{stars: pre-main sequence ---  stars: kinematics and dynamics --- stars: individual: T Cha --- protoplanetary disks --- techniques: interferometry}
   \authorrunning{Hu\'elamo et al.}
   \titlerunning{ALMA observations of  T Cha} 
   \maketitle
%

\section{Introduction}

\object{T Chamaeleontis} (\object{T Cha}) is a young ($\sim$7$\pm$5\,Myr) nearby (108\,pc) 
T~Tauri star in the $\epsilon$-Cha association, surrounded by a transition disk \citep[][]{Alcala1993,Brown2007,Torres2008,Murphy2013}.
There is evidence of a dust gap within the disk, and  a yet unconfirmed 
substellar companion inside the gap \citep[see][]{Huelamo2011,Olofsson2013}.
If confirmed, the disk around T Cha can give us important clues about the physical
conditions for substellar formation at early evolutionary phases.

T Cha is surrounded by a very narrow inner disk that extends from 0.13 to 0.17\,AU \citep{Olofsson2013}, 
and an outer  disk whose main  properties have been  inferred from the modeling of its spectral energy
distribution \citep[SED, e.g.][]{Brown2007}.
\citet[][]{Cieza2011} showed that the models are highly degenerate and can fit the SED of T Cha
equally well either with a very compact outer dust disk (a few AUs wide) or a
much larger but more 'tenuous' disk, with a very steep surface density profile.
Both family of models  suggest a very peculiar outer disk with little or no dust beyond $\sim$ 40\,AU. 
On the other hand, the cold gas in the T Cha outer disk has been studied by \citet[][]{Sacco2014}. Their 
spatially unresolved  observations  suggest  the presence of a gaseous disk with an outer radius of 
R$_{\rm CO}$$\sim$80\,AU in  Keplerian rotation.

Overall, the disk around T~Cha shows properties similar to the so-called 'faint' disks, characterized 
by  weak  millimeter continuum emission
that can be result of different properties or processes \citep[e.g.][]{Pietu2014}.
In the case of T Cha there is evidence of dust clearing, grain growth,  and a  high disk inclination \citep[][]{Brown2007, Pascucci2009}.

In this work we present high quality observations of T Cha obtained with
the Atacama Large Millimeter Array (ALMA), which have allowed us to spatially 
resolve the outer disk around T~Cha for the first time.  ALMA has allowed us to derive basic parameters of the 
outer disk,  to break the degeneracy of radiative transfer models based on SED fitting, and to understand if it is 
peculiar in comparison with other circumstellar disks.


\section{Observations}

The observations were performed on 2012 July 01, 26 and November 03 at Band 7, as part of the ALMA Cycle~0 program 2011.0.00921.S.  The field of view was $\sim18''$. A total of three data sets were collected, using between 18 and 23 antennas of $12\,{\rm m}$ diameter and accounting for 6 hours of total integration time including overheads and calibration.   Weather conditions were good and stable,  with an average precipitable water vapor of 0.7 mm. The system temperature varied from 150 to $250\,{\rm K}$. 

The correlator was set to four spectral windows in dual polarization mode, centered at 345.796 GHz (${\rm CO}(3$--$2)$), 
342.883 GHz (${\rm CS}(7$--$6)$), 332.505 GHz (${\rm SO_{2}} (4(3,1)$--$3(2,2))$), and 330.588 GHz (${\rm ^{13}CO}(3$--$2)$). The effective bandwidth used was $468.75{\rm MHz}$, providing a velocity resolution of  $\sim$ 0.11 km\,s$^{-1}$ after Hanning smoothing. 

The ALMA calibration includes simultaneous observations of the $183\,{\rm GHz}$ water line with water vapor radiometers, which measure the water column in the antenna beam, later used to reduce the atmospheric phase noise.   Amplitude calibration was done using Juno and Titan,  and quasars J$1256$$-057$ 
and J$1147$$-6753$ were used to calibrate the bandpass and the complex gain fluctuations respectively.
Data reduction was performed using version 4.1 of the Common Astronomy Software Applications package (CASA). We applied self-calibration using the continuum and we used the task CLEAN for imaging the self-calibrated visibilities. The continuum image was produced by combining all of the line-free 
channels using uniform weighting (synthesized beam $0.52\arcsec \times 0.34\arcsec$, ${\rm P.A.}\!\sim\!\!29^\circ$; rms = 0.7 mJy beam$^{-1}$). For the ${\rm CO}(3$--$2)$ line we used Briggs weighting (beam $0.64\arcsec \times 0.48\arcsec$, ${\rm P.A.}\!\sim\!\!31^\circ$; rms per channel = 9 mJy 
beam$^{-1}$), and for the rest of the lines we used natural weighting, providing a synthesized beam $\sim$$0.8\arcsec \times 0.6\arcsec$, ${\rm P.A.}\!\sim\!\!25^\circ$ and an rms per channel of 11~mJy beam$^{-1}$ for the ${\rm ^{13}CO}$ and 7~mJy beam$^{-1}$ for the ${\rm CS}$ and the ${\rm SO_{2}}$.

\section{Results and discussion \label{Results}}

\subsection{Molecular emission line detections\label{Line}}

Molecular line emission  is detected for the transitions ${\rm CO}(3$--$2)$, ${\rm ^{13}CO}(3$--$2)$, and ${\rm CS}(7$--$6)$. All of them are spatially resolved for the first time for T~Cha.  No detection was found for 
${\rm SO_{2}} (4(3,1)$--$3(2,2))$, with a 3$\sigma$ upper limit of 20 mJy.

\begin{figure*}[!ht]
\includegraphics[angle=0,scale=0.6]{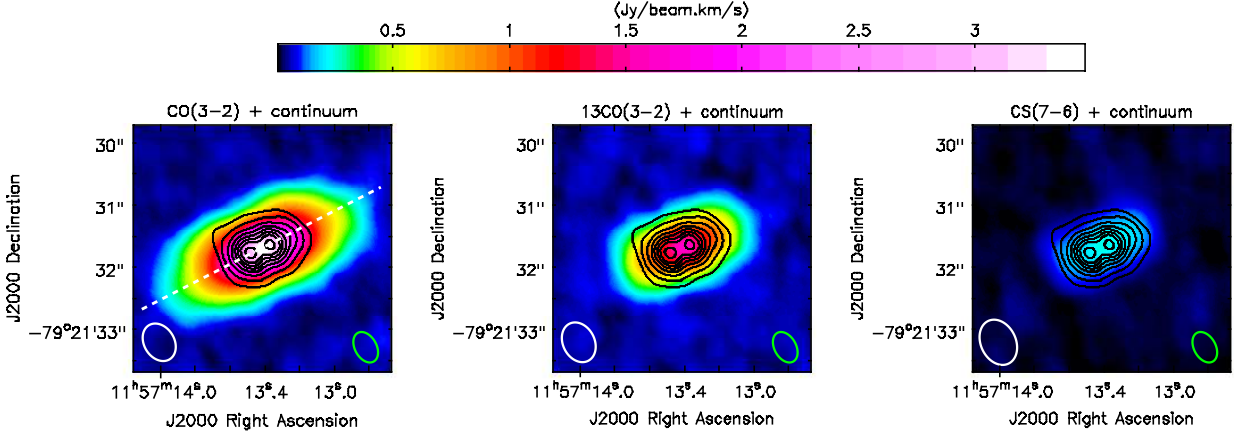}
\caption{Integrated emission maps of the ${\rm CO}(3$--$2)$, ${\rm ^{13}CO}(3$--$2)$, and the ${\rm CS}(7$--$6)$ transitions (from left to right). The black contours represent the continuum emission at 850~$\mu$m at 5, 15, 30, 45, 60, 75, 90, and 110~$\sigma$ where 1~$\sigma$ is 0.7 mJy beam$^{-1}$. 
We detect two emission bumps separated by 40\,AU and an outer dust radius of 79\,AU. The white ellipses are the synthesized beams for the spectral emission lines and the green ellipse is the synthesized beam for the continuum map. The white dashed line in the left panel represents the axis where the 
position-velocity diagram in Figure~\ref{PV2} has been obtained.}\label{MOM0}
\end{figure*}

\begin{figure*}[!ht]
\includegraphics[angle=0,scale=0.6]{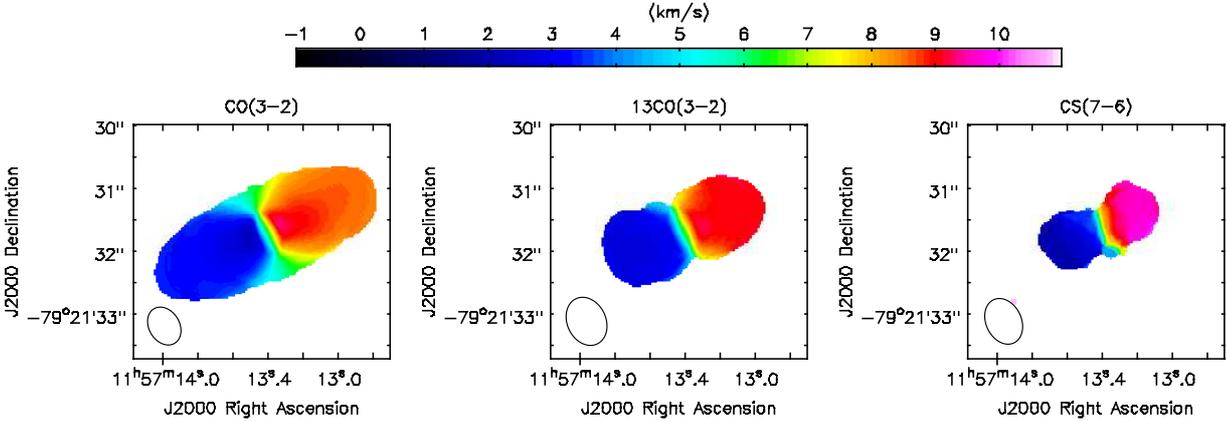}
\caption{Intensity-weighted mean velocity maps (first-order moment, 2$\sigma$ cut for ${\rm CO}(3$--$2)$  and  ${\rm^{13} CO}(3$--$2)$, and 1.5$\sigma$ cut for ${\rm CS}(7$--$6)$). }\label{MOM1}
\end{figure*}

Figure~\ref{MOM0} shows the integrated emission maps of the three detected molecules while Figure~\ref{MOM1} displays the intensity-weighted mean velocity maps. The ${\rm CO}(3$--$2)$ emission is spatially well resolved along the major and the minor axis. The integrated intensity averaged over the whole emission area above 3$\sigma$  is $12.46$ $\pm$ $0.11$ Jy km\,s$^{-1}$, in agreement with the value reported by \citet{Sacco2014} from the fit with a Keplerian disk model profile to single-dish observations. Considering contours above 3$\sigma$, the outer radius extends to 2.1$"$ which, after deconvolution with the synthesized beam, corresponds to an outer radius $R_{\rm CO}\!\sim230\,{\rm AU}$. 
The inclination ($i$) and the position angle ($PA$) of the gaseous disk have been estimated by fitting a Gaussian to the ${\rm CO}(3$--$2)$ integrated emission 
map, providing values of $i = 67\pm5\degree$ and $PA=113\pm6\degree$.

${\rm ^{13}CO}(3$--$2)$  emission line shows an integrated intensity of $4.31$$\pm$$0.07$ Jy km\,s$^{-1}$ (above 3$\sigma$ contour).  The region of emission is spatially resolved, with a radius $R_{\rm ^{13}CO}\!\sim170\,{\rm AU}$ after deconvolution with the synthesized beam.  

The radial intensity profiles of the CO(3-2) and $^{13}$CO(3-2) transitions are displayed in Figure~\ref{model_profiles}.  
They have been computed using slices along the semi-major axis of the disk (dashed white line in Fig.1, left panel) in the gas emission maps. 
We do not see a significant difference of the profiles at both sides of the disk (NE and SW) suggesting a symmetric distribution of the gas.

Gas emission from ${\rm CS}(7$--$6)$ is also spatially resolved and shows a deconvolved radius of $R_{\rm CS}\!\sim100 {\rm AU}$.  
The  integrated intensity emission above 3$\sigma$ is  $0.54$ $\pm$ $0.08$ Jy\,km\,s$^{-1}$. To our knowledge, this is the first time that such a high transition of the CS molecule is spatially resolved in a disk around a late-type star \citep[the first detection in a Herbig star has been recently reported by][]{plas2014}. Sulfur-bearing molecules (like CS and SO$_{2}$) studies are scarce (e.g. \citealt{Dutrey2011,Guilloteau2012}), but interesting since these type of molecules are present in comets (e.g. Hale-Bopp; \citealt{Ikeda2002},  and Shoemaker-Levy/9; \citealt{Matthews2002}) and their presence in the inner part of the disks offers the possibility of studying the chemical composition of the planet-forming regions.  So far, only a few CS detections at lower frequency  transitions have been reported in disks surrounding K5 to M0 stars \citep{Dutrey2011,Kastner2013}.  Here we show for the first time a spatially resolved detection of the CS(7--6) transition around a K0 star.

The gas detected in the disk surrounding T Cha shows a velocity profile consistent with a Keplerian rotation pattern,  which can be seen in all the molecular transitions observed (see Figure~\ref{MOM1}).   ${\rm CO}(3$--$2)$ emission is detected at velocities between -5.0 to 16.5\,km\,s$^{-1}$, ${\rm ^{13}CO}(3$--$2)$ between -3.0 and 15.0 km\,s$^{-1}$ and the ${\rm CS}(7$--$6)$ between 0.0 to 11.0 km\,s$^{-1}$. In Fig~\ref{PV2} we have represented a position-velocity diagram along the major axis of the gaseous disk traced by ${\rm CO}(3$--$2)$.  The systemic velocity is 5.95$\pm$0.22 km\,s$^{-1}$.  In that plot an absorption feature at velocities near 4.7\,km\,s$^{-1}$ can be seen, which is very likely produced by a foreground molecular cloud in the same line of sight as T~Cha \citep{Nehme2008}.
By comparing the position-velocity diagram with a Keplerian velocity profile (Figure ~\ref{PV2}), 
we show that the emission is compatible with Keplerian rotation around a 1.3 - 1.7 M$_{\odot}$ 
object in a disk inclined at 67 degrees. This mass range is in good agreement with estimations from evolutionary tracks 
for an age of $\sim$10\,Myr \citep[e.g.][]{Schisano2009,Murphy2013}.


\subsection{ Continuum emission at $850\,\mu{\rm m}$ \label{Continuum}}

Continuum emission at $850\,\mu{\rm m}$ is detected centered at the position R.A.(J2000) = 11$^{h}$57$^{m}$13$\fs$42, Dec(J2000) =$-$79$\degr$21$\arcmin$31$\arcsec$696. The flux density integrated over the disk structure above 5$\sigma$ is  $S_{850\mu m}$ = 198 $\pm$ 4 mJy.  

Figure~\ref{MOM0} shows the dusty outer disk around T~Cha represented by black contours. The disk is spatially resolved in its major axis with a projected diameter of $1.50''$ (measured at the 5$\sigma$ contour level), which corresponds to an outer dust disk radius of  $R_{\rm out}\!\sim80\,{\rm AU}$ after deconvolution with the synthesized beam, and adopting a distance of $108\,{\rm pc}$.   Two local peaks are observed at a projected separation of $0.37"$ (40 AU at 108\,pc),  which is close to the beam size, and suggest the presence of a gap in the inner regions of the disk as predicted by SED modeling.


In Figure~\ref{model_profiles} we have represented the continuum radial intensity profiles at both sides of the disk, together with the average profile, including only datapoints above 5-$\sigma$. As in the case of the gas molecules, 
they have been computed using slices along the semi-major axis of the disk (dashed white line in Fig.1, left panel) in the continuum emission maps.  
We can see a significant  difference between the profiles at both sides,  with the NW side being slightly larger than the SE one, suggesting asymmetries within the dusty disk. Finally, the $i $ and $PA$ values that we derive using the continuum emission at 
850~$\mu$m are similar to those estimated with the ${\rm CO}(3$--$2)$ observations.

\subsection{Comparison with radiative transfer models}

Our ALMA observations reveal that T Cha is surrounded by a compact dusty disk with a sharp outer edge at $\sim$80\,AU 
and a larger gaseous disk with an outer radius of $\sim$230\,AU.  This trend, a compact dust disk with a larger and more 
diffuse gaseous disk, has already been observed in a significant number of circumstellar disks 
\citep[e.g.][]{Isella2007,Hughes2008,Andrews2012,deGregorio2013,Pietu2014}.

\citet{Cieza2011} used the radiative transfer code MCFOST \citep{Pinte2006,pinte2009} to model
the SED of T~Cha. They showed that there are two families of dust disk models
that can reproduce equally well  the SED:  
very small disks (a few AUs width) or  much larger ($R_{\rm out}$$\sim$300\,AU) but with 
a very steep surface density profile (with an  exponent of $\alpha$ $\leq$ --2, for a power-law prescription). 
The large degeneracy between these two disk parameters, $R_{\rm out}$ and $\alpha$, did not 
allow them to choose between these two scenarios. 
\citet{Olofsson2013} used MCFOST to  fit the SED together with near-IR interferometric data. They fixed
$R_{\rm out}$=25\,AU,  and $\alpha$=-1, and found a best fit model with a narrow dust outer disk ($R_{\rm in}$=12$\pm2$\,AU).

Our ALMA data shows that models with a power-law surface density, like the ones used in these two works, 
cannot reproduce both the CO and continuum data. We nevertheless first consider this type of models 
to discuss the ALMA data in the context of previous results.

We have modeled the SED of T Cha  with MCFOST, being the
starting point the model grid presented by
\cite{Cieza2011} and refined by \cite{Olofsson2013}.
Basically, the disk is composed of 2 sub-disks: an inner and an outer
disk with a density structure defined by a power law surface density profile with exponent $\alpha$,
$\Sigma (r)\, = \Sigma_{0}\,(r/r_0)^\alpha$, and a scale height of 
$ h(r)\,=\,h_{0} (r/r_{0})^\beta$, with $\beta$ being the disk flaring index, and $h_0$ the 
scale height at a reference radius $r_0$ = 50\,AU.
Each disk extends from an inner radius, $R_{\rm in}$ to an outer radius, $R_{\rm out}$.
The grain size distribution in each disk is defined by 
$\mathrm{d}n(a) \propto a^p\mathrm{d}a$,  between $a_\mathrm{min}$ and $a_\mathrm{max}$.
The temperature structure in the disk is calculated by considering the dust
opacity only. The dust properties are computed assuming the Mie theory.
Finally, the total dust mass in the disk (including all grain sizes) is represented by $M_{\rm dust}$.

\begin{figure}[!t]
\includegraphics[angle=0,scale=0.48]{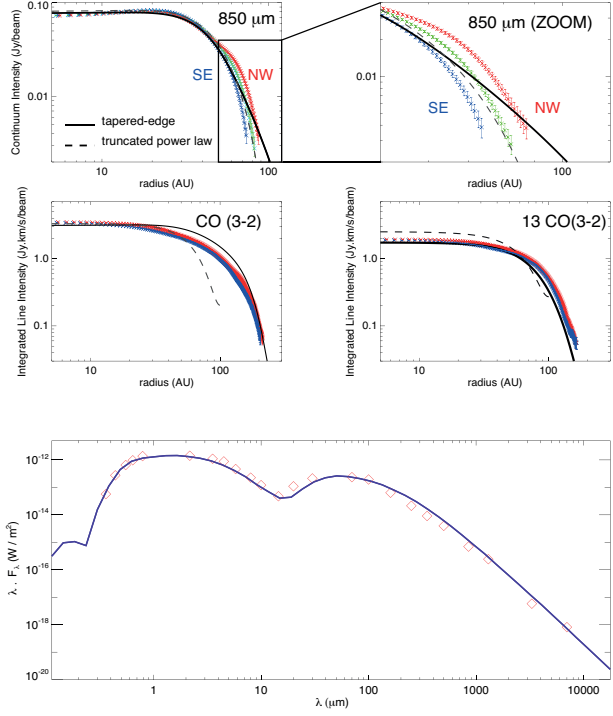}
\caption{Top and middle panels: Continuum and gas radial intensity profiles. The blue (SE) and red (NW) data points are the ALMA data (over 5-$\sigma$) at both sides of the disk. In the case  of the continuum we have also included the average profile (green data). The solid black lines show the best model using a tapered edge prescription for the surface density, while the dashed lines show the best model
using a truncated power law. The upper panels show the continuum data, including a zoom of the outer regions, while the middle panels include the CO profiles. Bottom panel: the observed SED and the fit from the tapered edge model presented here.}\label{model_profiles}
\end{figure}
\begin{figure}[!h]
\includegraphics[angle=0,scale=0.42]{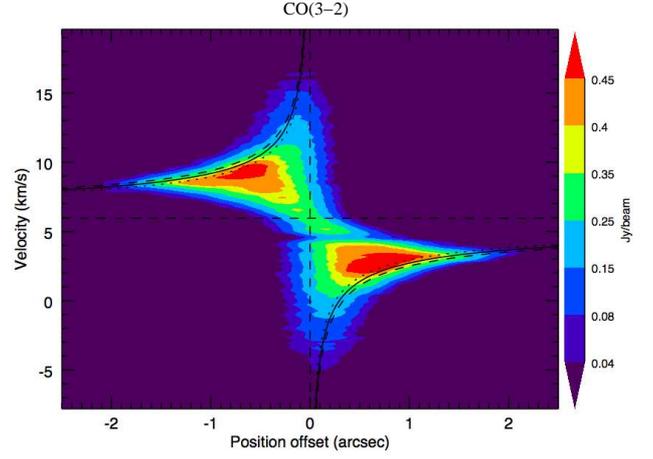}
\caption{Position-velocity diagram along the disk major axis (see exact axis represented in Fig~\ref{MOM0} left panel). The dotted, solid, and dashed black curves represent the best fits to the data, that correspond to a Keplerian velocity profile for a systemic velocity of 5.95 km s$^{-1}$, an inclination of $i(\degree)$\,=\,67, and a central mass of 1.3, 1.5, and 1.7 ${\rm M_{\odot}}$, respectively.}\label{PV2}
\end{figure}

To calculate the CO channel maps and surface brightness distribution,
we assume a constant gas-to-dust mass ratio of 100 throughout the disk
(both radially and vertically). We adopted a standard CO abundance with respect
to H$_2$ ($10^{-4}$), set constant through the disk where $T_{\rm dust}$$>$ 20\,K and equal to
zero where $T_{\rm dust}$$<$ 20\,K to mimic the effect of CO freeze out.
The $^{12}$CO/$^{13}$CO ratio is set to 76. The level populations are calculated assuming 
$T_{\rm gas}$ = $T_{\rm dust}$ at each point in the disk . The radial and vertical temperature profiles
and the radiation field estimated by the Monte Carlo simulation are used to
calculate level populations for the CO molecule and to produce the SED,
continuum images, and line emission surface brightness profiles, as well as
kinematics with a ray-tracing method. The kinematics are calculated assuming
the disk is in pure Keplerian rotation.

We have adopted the inner disk parameters from \citet{Olofsson2013}.
For the outer disk, we have fixed two parameters obtained from the 
ALMA data, $R_{\rm out}$ and $i$, selecting the grid values closer to the
ALMA measurments (80\,AU and 68\degree, respectively).
We have explored $R_{\rm in}$, $\beta$, $h_\circ$@50AU, $a_{\rm max}$, $\alpha$, and $M_{\rm dust}$,
using the same parameter range shown in \citet{Cieza2011}.
For the SED, we have fitted the same  observational dataset displayed in that work.
The adopted stellar parameters are $T_{\rm eff}$=5400\,K, $A_{\rm v}$=1.5, and 
$d$=108\,pc \citep{Torres2008,Schisano2009}. The best disk model, that is, the one with the minimum $\chi$$^2$, 
provides parameters of $\alpha$= -2.5, $\beta$=1.07, $h_0$@50AU\,=\,6\,AU, $R_{\rm in}$ = 19\,AU, $a_{\rm max}$ = 1000,  
and a disk dust mass of $M_{\rm dust}$ $\sim$1$\times$10$^{-5}$\,M$_{\odot}$.

\begin{figure}[t!]
\includegraphics[angle=0,scale=0.4]{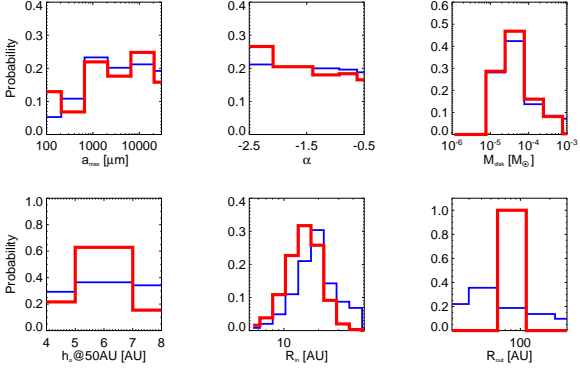}
\caption{Bayesian probability distributions of the disk parameters of T Cha. The blue lines show the results without fixing any parameter,  while the red lines show the results after fixing the outer disk radius and the disk inclination.}\label{bayesian}
\end{figure}

Since SED modeling is highly degenerate, the best-fit model is unlikely to be a unique solution. Therefore, we have performed 
a Bayesian analysis  to  estimate the validity range for each of the explored parameters \citep{Press1992,Pinte2007}.
The result is displayed in Figure~\ref{bayesian}, where we show the  Bayesian probability distributions for the  different disk parameters.  
While $M_{\rm dust}$, $R_{\rm in}$ and $h_{\rm 0}$ seem well constrained, this is not the case for  $\alpha$: it shows a local peak at $\alpha= -2.5$ 
but a relatively flat distribution. We conclude that, even fixing $R_{\rm out}$, $\alpha$ remains unconstrained by the SED modeling.

With our ALMA observations we have partially broken the ($\alpha$, $R_{\rm out}$) degeneracy commonly encountered with SED fitting by measuring $R_{\rm out}$. 
The resolution reached by our observations does not allow us to constrain accurately the surface density profile.  But with R$_{\rm in}$ $\sim$ 20\,AU 
and a disk width of  $\sim$60 AU, we exclude surface density profile shallower than -1.
According to the observed $R_{\rm out}$, we can also discard the family of models with very narrow dusty rings, and the extreme case of a very large disk 
($R_{\rm out}$ $\sim$300 AU) with $\alpha$$\sim$-3.

If we take the gas emission into account, the model fails to fit simultaneously the gas and dust 
profiles (see Figure~\ref{model_profiles}), as already observed in  other spatially resolved circumstellar disks 
\citep[e.g.][]{Hughes2008}.  As discussed by  the authors, a power law density
profile cannot reproduce the different extent of the gas and dust emission observed in circumstellar disks while  
a tapered edge model, in which the surface density falls off gradually, can in principle reconcile 
the observed profiles. We have therefore run the MCFOST code, but using a tapered exponential edge in 
the surface density distribution. In this case, the density profile is represented by a function defined by the 
characteristic radius, $R_{\rm c}$ (the radius out of which the brightness drops towards zero) and
$\gamma$, the surface density index:
$\Sigma(r) = \Sigma_0 (r/r_0)^{-\gamma} \exp\left( -(r/r_0)^{2-\gamma}\right)$.  

For this model, we have also adopted the inner disk parameters from \citet[][]{Olofsson2013}, and varied only the 
prescription for the outer disk. We have fixed the disk inclination to 68$^{\circ}$, the dust mass to $M_{\rm dust}$= 9$\times$ 10$^{-5}$\,M$_{\odot}$  (de Gregorio-Monsalvo et al., in prep.), and we have explored  the rest of the disk parameters:  we have sampled a range of $\gamma$ values 
 between  0 and 2, $R_{\rm c}$ between 40 and 100\,AU, $R_{\rm in}$ between 12 and 30\,AU,  the scale height  at 50\,AU
 ($h_{\rm 0}$@50AU) between 3 and 6\,AU,  and the flaring index, $\beta$, between 1.0 and 1.1.
 Finally, to take into account the difference in brightness between both sides of the disk in the continuum,
 we explore models that pass between the two observed profiles.

The result is displayed in Figure~\ref{model_profiles} where we show the best model that can
fit simultaneously the  two disk components and the observed SED.
The model shows $\gamma$ = 0.5  and  $R_{\rm c}$ = 50\,AU for both the gas and the dust.
We also derive these parameters: $R_{\rm in}$= 20\,AU,  $h_{\rm 0}$@50\,AU = 4\,AU,  and $\beta$ = 1.0. 
This  model is consistent with having the gas and the dust well mixed and mainly located 
at a radius smaller than 50\,AU, as suggested by \citet{Cieza2011} based on the 
steep drop of the SED at sub-mm wavelengths.

Figure~\ref{model_profiles} shows that our best model does not perfectly fit the CO line profiles, which can be related 
either  with the underlying chemistry (we assumed ISM abundances and very simple CO freeze-out) or with the model prescriptions. In fact, tapered-edge models sometimes fail to reproduce simultaneously the observed dust and gas profiles obtained from  very high spatial resolution and sensitivity observations 
\citep[see][]{Andrews2012,deGregorio2013}, and suggests that other  processes like e.g.  grain growth and radial migration should be taken into account.  
Given that the disk is barely resolved in our observations, we expect future, higher spatial resolution observations to provide stronger constraints on the relative location
of the  gas and dust, and the departure from point-symmetry.

\section{Conclusions \label{Conclusion}}

High spatial resolution and high sensitivity ALMA observations have allowed us to spatially resolve the outer disk around 
the young and isolated object T Cha. The target is surrounded by a compact dusty disk and 
a $\sim$3 times larger gaseous disk.  Our main results can be summarized as follows:

\begin{itemize}

\item We have spatially resolved the gaseous disk of T Cha in three different molecular emission lines: CO(3--2), $^{13}$CO(3-2) and CS(7--8). Using the  CO(3-2) image we derive an outer radius of $R_{\rm gas,out}\sim$230\,AU, an inclination of $i (\degree)$\,=\,67$\pm$5, and a position angle of $PA (\degree)$\,=\,113$\pm$6.  
The line intensity profiles are similar at both sides of the disk in the CO molecules, consistent with a uniform distribution of the gas.

\item The disk around T Cha is in Keplerian rotation, and the estimated dynamical mass of the central object, 
$M_{*}$\,=\,1.5$\pm$0.2\,M$_{\odot}$, is in good agreement with previous estimations based on evolutionary tracks.

\item The dusty disk is resolved in the continuum observations at 850\,$\mu$m and it shows a similar $i$ and $PA$ to the gaseous disk. 
The continuum intensity profile displays two emission bumps separated by 40\,AU, suggesting the presence of an inner dust gap as predicted by SED modeling, and  
an outer radius of $\sim$80\,AU. The  profiles are different at both sides of the disk, which points towards asymmetries in the dust distribution. 
These data allows us to rule out both the very small and large $R_{\rm out}$ families of SED models.

\item Radiative transfer models including a truncated power law prescription for the surface density profile  cannot reproduce simultaneously the  gas and dust profiles. We can fit both components simultaneously using a tapered-edge model prescription for the surface density.
The best model provides values of $\gamma$= 0.5, and $R_{\rm c}$= 50\,AU, which is consistent with having most of the 
disk mass within the inner 50\,AU.

\end{itemize}

\begin{acknowledgements}
 This paper makes use of the following ALMA data: ADS/JAO.ALMA\#2011.0.00921.S. ALMA is a partnership of ESO (representing its member states), NSF (USA) and NINS (Japan), together with NRC (Canada) and NSC and ASIAA (Taiwan), in cooperation with the Republic of Chile. The Joint ALMA Observatory is operated by ESO, AUI/NRAO and NAOJ. The NRAO is a facility of the National Science Foundation operated under cooperative agreement by Associated Universities, Inc. This research has been funded by Spanish grants AYA2010-21161-C02-02 and AYA2012-38897-C02-01. IdG and EM acknowledge support from MICINN (Spain) AYA2011-30228-C03 grant (including FEDER funds).   

\end{acknowledgements}

\bibliographystyle{aa}
\bibliography{almatcha}

\begin{thebibliography}{26}
\expandafter\ifx\csname natexlab\endcsname\relax\def\natexlab#1{#1}\fi

\bibitem[{{Alcala} {et~al.}(1993){Alcala}, {Covino}, {Franchini}, {Krautter},
  {Terranegra}, \& {Wichmann}}]{Alcala1993}
{Alcala}, J.~M., {Covino}, E., {Franchini}, M., {et~al.} 1993, \aap, 272, 225

\bibitem[{{Andrews} {et~al.}(2012){Andrews}, {Wilner}, {Hughes}, {Qi},
  {Rosenfeld}, {{\"O}berg}, {Birnstiel}, {Espaillat}, {Cieza}, {Williams},
  {Lin}, \& {Ho}}]{Andrews2012}
{Andrews}, S.~M., {Wilner}, D.~J., {Hughes}, A.~M., {et~al.} 2012, \apj, 744,
  162

\bibitem[{{Brown} {et~al.}(2007){Brown}, {Blake}, {Dullemond}, {Mer{\'{\i}}n},
  {Augereau}, {Boogert}, {Evans}, {Geers}, {Lahuis}, {Kessler-Silacci},
  {Pontoppidan}, \& {van Dishoeck}}]{Brown2007}
{Brown}, J.~M., {Blake}, G.~A., {Dullemond}, C.~P., {et~al.} 2007, \apjl, 664,
  L107

\bibitem[{{Cieza} {et~al.}(2011){Cieza}, {Olofsson}, {Harvey}, {Pinte},
  {Mer{\'{\i}}n}, {Augereau}, {Evans}, {Najita}, {Henning}, \&
  {M{\'e}nard}}]{Cieza2011}
{Cieza}, L.~A., {Olofsson}, J., {Harvey}, P.~M., {et~al.} 2011, \apjl, 741, L25

\bibitem[{{de Gregorio-Monsalvo} {et~al.}(2013){de Gregorio-Monsalvo},
  {M{\'e}nard}, {Dent}, {Pinte}, {L{\'o}pez}, {Klaassen}, {Hales},
  {Cort{\'e}s}, {Rawlings}, {Tachihara}, {Testi}, {Takahashi}, {Chapillon},
  {Mathews}, {Juhasz}, {Akiyama}, {Higuchi}, {Saito}, {Nyman}, {Phillips},
  {Rod{\'o}n}, {Corder}, \& {Van Kempen}}]{deGregorio2013}
{de Gregorio-Monsalvo}, I., {M{\'e}nard}, F., {Dent}, W., {et~al.} 2013, \aap,
  557, A133

\bibitem[{{Dutrey} {et~al.}(2011){Dutrey}, {Wakelam}, {Boehler}, {Guilloteau},
  {Hersant}, {Semenov}, {Chapillon}, {Henning}, {Pi{\'e}tu}, {Launhardt},
  {Gueth}, \& {Schreyer}}]{Dutrey2011}
{Dutrey}, A., {Wakelam}, V., {Boehler}, Y., {et~al.} 2011, \aap, 535, A104

\bibitem[{{Guilloteau} {et~al.}(2012){Guilloteau}, {Dutrey}, {Wakelam},
  {Hersant}, {Semenov}, {Chapillon}, {Henning}, \&
  {Pi{\'e}tu}}]{Guilloteau2012}
{Guilloteau}, S., {Dutrey}, A., {Wakelam}, V., {et~al.} 2012, \aap, 548, A70

\bibitem[{{Hu{\'e}lamo} {et~al.}(2011){Hu{\'e}lamo}, {Lacour}, {Tuthill},
  {Ireland}, {Kraus}, \& {Chauvin}}]{Huelamo2011}
{Hu{\'e}lamo}, N., {Lacour}, S., {Tuthill}, P., {et~al.} 2011, \aap, 528, L7

\bibitem[{{Hughes} {et~al.}(2008){Hughes}, {Wilner}, {Qi}, \&
  {Hogerheijde}}]{Hughes2008}
{Hughes}, A.~M., {Wilner}, D.~J., {Qi}, C., \& {Hogerheijde}, M.~R. 2008, \apj,
  678, 1119

\bibitem[{{Ikeda} {et~al.}(2002){Ikeda}, {Kawaguchi}, {Takakuwa}, {Sakamoto},
  {Sunada}, \& {Fuse}}]{Ikeda2002}
{Ikeda}, M., {Kawaguchi}, K., {Takakuwa}, S., {et~al.} 2002, \aap, 390, 363

\bibitem[{{Isella} {et~al.}(2007){Isella}, {Testi}, {Natta}, {Neri}, {Wilner},
  \& {Qi}}]{Isella2007}
{Isella}, A., {Testi}, L., {Natta}, A., {et~al.} 2007, \aap, 469, 213

\bibitem[{{Kastner} {et~al.}(2013){Kastner}, {Punzi}, {Rodriguez}, {Sacco},
  {Hily-Blant}, {Forveille}, \& {Zuckerman}}]{Kastner2013}
{Kastner}, J., {Punzi}, K., {Rodriguez}, D., {et~al.} 2013, in Protostars and
  Planets VI Posters, 22

\bibitem[{{Matthews} {et~al.}(2002){Matthews}, {Marten}, {Moreno}, \&
  {Owen}}]{Matthews2002}
{Matthews}, H.~E., {Marten}, A., {Moreno}, R., \& {Owen}, T. 2002, \apj, 580,
  598

\bibitem[{{Murphy} {et~al.}(2013){Murphy}, {Lawson}, \& {Bessell}}]{Murphy2013}
{Murphy}, S.~J., {Lawson}, W.~A., \& {Bessell}, M.~S. 2013, \mnras, 435, 1325

\bibitem[{{Nehm{\'e}} {et~al.}(2008){Nehm{\'e}}, {Gry}, {Boulanger}, {Le
  Bourlot}, {Pineau Des For{\^e}ts}, \& {Falgarone}}]{Nehme2008}
{Nehm{\'e}}, C., {Gry}, C., {Boulanger}, F., {et~al.} 2008, \aap, 483, 471

\bibitem[{{Olofsson} {et~al.}(2013){Olofsson}, {Benisty}, {Le Bouquin},
  {Berger}, {Lacour}, {M{\'e}nard}, {Henning}, {Crida}, {Burtscher}, {Meeus},
  {Ratzka}, {Pinte}, {Augereau}, {Malbet}, {Lazareff}, \&
  {Traub}}]{Olofsson2013}
{Olofsson}, J., {Benisty}, M., {Le Bouquin}, J.-B., {et~al.} 2013, \aap, 552,
  A4

\bibitem[{{Pascucci} \& {Sterzik}(2009)}]{Pascucci2009}
{Pascucci}, I. \& {Sterzik}, M. 2009, \apj, 702, 724

\bibitem[{{Pi{\'e}tu} {et~al.}(2014){Pi{\'e}tu}, {Guilloteau}, {Di Folco},
  {Dutrey}, \& {Boehler}}]{Pietu2014}
{Pi{\'e}tu}, V., {Guilloteau}, S., {Di Folco}, E., {Dutrey}, A., \& {Boehler},
  Y. 2014, \aap, 564, A95

\bibitem[{{Pinte} {et~al.}(2007){Pinte}, {Fouchet}, {M{\'e}nard}, {Gonzalez},
  \& {Duch{\^e}ne}}]{Pinte2007}
{Pinte}, C., {Fouchet}, L., {M{\'e}nard}, F., {Gonzalez}, J.-F., \&
  {Duch{\^e}ne}, G. 2007, \aap, 469, 963

\bibitem[{{Pinte} {et~al.}(2009){Pinte}, {Harries}, {Min}, {Watson},
  {Dullemond}, {Woitke}, {M{\'e}nard}, \& {Dur{\'a}n-Rojas}}]{pinte2009}
{Pinte}, C., {Harries}, T.~J., {Min}, M., {et~al.} 2009, \aap, 498, 967

\bibitem[{{Pinte} {et~al.}(2006){Pinte}, {M{\'e}nard}, {Duch{\^e}ne}, \&
  {Bastien}}]{Pinte2006}
{Pinte}, C., {M{\'e}nard}, F., {Duch{\^e}ne}, G., \& {Bastien}, P. 2006, \aap,
  459, 797

\bibitem[{{Press} {et~al.}(1992){Press}, {Teukolsky}, {Vetterling}, \&
  {Flannery}}]{Press1992}
{Press}, W.~H., {Teukolsky}, S.~A., {Vetterling}, W.~T., \& {Flannery}, B.~P.
  1992, {Numerical recipes in FORTRAN. The art of scientific computing}
  (Cambridge: University Press, |c1992, 2nd ed.)

\bibitem[{{Sacco} {et~al.}(2014){Sacco}, {Kastner}, {Forveille}, {Principe},
  {Montez}, {Zuckerman}, \& {Hily-Blant}}]{Sacco2014}
{Sacco}, G.~G., {Kastner}, J.~H., {Forveille}, T., {et~al.} 2014, \aap, 561,
  A42

\bibitem[{{Schisano} {et~al.}(2009){Schisano}, {Covino}, {Alcal{\'a}},
  {Esposito}, {Gandolfi}, \& {Guenther}}]{Schisano2009}
{Schisano}, E., {Covino}, E., {Alcal{\'a}}, J.~M., {et~al.} 2009, \aap, 501,
  1013

\bibitem[{{Torres} {et~al.}(2008){Torres}, {Quast}, {Melo}, \&
  {Sterzik}}]{Torres2008}
{Torres}, C.~A.~O., {Quast}, G.~R., {Melo}, C.~H.~F., \& {Sterzik}, M.~F. 2008,
  {Young Nearby Loose Associations}, ed. {Reipurth, B.}, 757--+

\bibitem[{{van der Plas} {et~al.}(2014){van der Plas}, {Casassus},
  {M{\'e}nard}, {Perez}, {Thi}, {Pinte}, \& {Christiaens}}]{plas2014}
{van der Plas}, G., {Casassus}, S., {M{\'e}nard}, F., {et~al.} 2014, \apjl,
  792, L25

\end{thebibliography}

\end{document}